\input{aipcheck}

\documentclass[
    ,final            
  ]
  {aipproc}

\layoutstyle{8x11double}

\usepackage{natbib}

\begin{document}

\title{The INTEGRAL view of Gamma-Ray Bursts}

\classification{95.85.Pw $\gamma$-ray; 98.70.Rz $\gamma$-ray sources; $\gamma$-ray bursts; 95.80.+p sky surveys; 95.55.Ka X- and $\gamma$-ray telescopes and instrumentation; 95.30.Gv Radiation mechanisms; polarization}
\keywords{$\gamma$-rays, GRB, surveys}

\author{P. Ubertini}{
  address={IASF-Roma, INAF, Via Fosso del Cavaliere, 100 - 00133 Roma, Italy}
}

\author{A. Corsi}{
  address={IASF-Roma, INAF, Via Fosso del Cavaliere, 100 - 00133 Roma, Italy}
,altaddress={Universit\`a di Roma ``Sapienza'', Piazzale Aldo Moro 2 - 00185 Roma, Italy}
}

\author{S. Foley}{
  address={UCD School of Physics, University College Dublin, Dublin 4, Ireland}
}

\author{S. McGlynn}{
  address={Department of Physics, Royal Institute of Technology (KTH), AlbaNova, SE-106 91 Stockholm, Sweden}
}

\author{G. De Cesare}{
  address={IASF-Roma, INAF, Via Fosso del Cavaliere, 100 - 00133 Roma, Italy}
}

\author{A. Bazzano}{
  address={IASF-Roma, INAF, Via Fosso del Cavaliere, 100 - 00133 Roma, Italy}
}

 
\renewcommand\XFMtitleblock{%
  \XFMtitle
  \let\XFMoldpar\par
  \def\par{\XFMoldpar\def\par{\space 
             on behalf of the IBIS Team\XFMoldpar}}%
   \XFMauthors
   \let\par\XFMoldpar
   \XFMaddresses
   \XFMabstract
   \vspace{5pt}%
   \XFMkeywords
   \XFMclassification
 }

\begin{abstract}
After more than six and half years in orbit, the ESA
space observatory INTEGRAL has provided new, 
exciting results in the soft gamma-ray energy range (from a few keV to a few  MeV). 
With the discovery of about 700 hard X-Ray sources, it 
has changed our previous view of a sky composed 
of peculiar and ``monster'' sources. The new high energy sky is in fact full of a large
variety of normal, very energetic emitters, characterized by
new accretion and acceleration processes (see also IBIS cat4 \citep{Bird:2009}). 
If compared to previous IBIS/ISGRI surveys it is clear that there is a continual increase in the
rate of discoveries of HMXB and AGN, including a variety of distant QSOs. This is basically due to
increased exposure away from the Galactic Plane, while the
percentage of sources without an identification has remained
constant. At the same time, about one GRB/month is 
detected and imaged by the two main gamma-ray instruments on board: IBIS and SPI.
 INTEGRAL, after six and half years of observations, has completed the Core
Programme phase and is now fully open to the scientific
community for Open Time and Key Programme observations, with AO7 recently announced by ESA.  

In this paper we review the major achievements of the INTEGRAL Observatory in the field of Gamma Ray Bursts.
\end{abstract}

\maketitle


\section{The INTEGRAL Observatory}
The ESA INTEGRAL (International Gamma-Ray Astrophysics Laboratory)
observatory was selected in June 1993 as the next medium-size
scientific mission within the ESA Horizon 2000 programme.
INTEGRAL \citep{Winkler:2003}, is
dedicated to the fine spectroscopy (2.5 keV FWHM @ 1 MeV) and fine
imaging (angular resolution: 12 arcmin FWHM) of celestial
gamma-ray sources in the energy range 15 keV to 10 MeV. While the
imaging is carried out by the imager IBIS \citep{Ubertini:2003},
the fine spectroscopy is performed by the spectrometer SPI
\citep{Vedrenne:2003} and coaxial monitoring capability in the
X-ray (3-35 keV) and optical (V-band, 550 nm) energy ranges is
provided by the JEM X and OMC instruments \citep{Lund:2003,Mass-Hesse:2003}. 
SPI, IBIS and Jem-X, the spectrometer,
imager and X-ray monitor, are based on the use of the coded aperture
mask technique, which is a key method used to provide images at energies
above tens of keV, where photon focusing becomes impossible using
standard grazing techniques. Moreover, coded masks optimise
background subtraction capability because of the opportunity to
observe the source and the off-source sky at the same time. This
is achieved simultaneously for all the sources present in the
telescope field of view. In fact, for any particular source
direction, the detector pixels are divided into two sets: those
capable of viewing the source and those for which the flux is
blocked by opaque tungsten mask elements. This very well
established technique is discussed in detail by \citep{Skinner:2003}
and is extremely effective in controlling the
systematic error in all telescope observations, working
remarkably well for weak extragalactic field sources as well as for crowded
galactic regions, such as our Galactic Center. The mission was
conceived from the beginning as an observatory led by ESA with
contributions from Russia (PROTON launcher) and NASA (Deep Space
Network ground station). INTEGRAL was launched from Baikonur on
October 17$^{th}$, 2002 and inserted into an highly eccentric
orbit (characterized by 9000 km perigee and 154000 km apogee). The
high perigee was used in order to provide long uninterrupted observations
with nearly constant background and away from the electron and
proton radiation belts. Scientific observations can then be
carried out while the satellite is above a nominal altitude of
60000 km while entering the radiation belts, and above 40000 km
while leaving them. This strategic choice ensures that about 90\% of the
time is used for scientific observations with a data rate of
realtime 108 kbps science telemetry being received from the ESA station
at Redu and NASA station at Goldstone. The data are received by
the INTEGRAL Mission Operation Centre (MOC) in Darmstadt (Germany)
and relayed to the Science Data Center (ISDC, \citep{Courvoisier:2003})
which provides the final consolidated data products to the observers
and later archives them for public use. The proprietary data become
public one year after their distribution to single observation PIs.

To date INTEGRAL has already carried out
more than 6 years of Core, Key and Guest Observer Programme
observations in the energy range from 5 keV up to 10 MeV with high angular 
resolution (few arcmin PSLA of the IBIS imager) and excellent spectral capability 
(2.5 keV spectral resolution of the SPI germanium spectrometer).

\section{Highlights from the Integral GRB sample}

\subsection{The 1st bursts in the sample}
\subsubsection{GRB 021125}
Just a few weeks after its launch, INTEGRAL started an in-orbit calibration observing Cyg X-1. On
November 25th, 2002, the satellite was set up for a special observation
with the PICsIT layer in the non-standard photon-by-photon mode \citep{DiCocco:2003}. Since 
PICsIT operates in an energy band where the background
rate is very high, to perform the calibration it was necessary to
limit the operative range below 500 keV and to strongly reduce
the telemetry allocation to the other instruments (SPI, JEM-X,
OMC were sending only housekeeping data to ground) and to
the ISGRI layer of IBIS. During this test, at 17:58:30 UTC a GRB occurred
in the partially coded field of view of IBIS (about $7.3^{\circ}$ off-axis)
and lasted about 24 s \citep{Bazzano:2002,Malaguti:2003}.
The burst was soon confirmed by the Inter Planetary Network
(IPN, \citep{Hurley:2002}). GRB 021125 was the first GRB detected by INTEGRAL in the
field of view of the IBIS imager. The sky coordinates reconstructed
with IBIS and SPI were in agreement with each other,
and consistent with the IPN error box.
The spectrum and light curve, obtained using independent methods,
yielded consistent results. GRB 021125 thus proved the capabilities of the instruments
on board the INTEGRAL satellite for GRB observations.

\subsubsection{GRB 030131}
Although not specifically designed as a
GRB-oriented mission, INTEGRAL has contributed to the rapid localization
of GRBs thanks to the INTEGRAL
Burst Alert System (IBAS, \citep{Mereghetti:2001}). This software,
running at the ISDC \citep{Courvoisier:2003}, is able to detect and localize GRBs
with a precision of a few arcminutes in a few seconds, and to
distribute their coordinates in near real time over the Internet.
In addition, the high sensitivity of the INTEGRAL instruments is particularly well-suited 
to study in detail the prompt $\gamma$-ray emission of the faintest bursts. 
On January 31 2003 at 07:38:49 UTC, GRB 030131 was
detected in the field of view of IBIS and SPI.
The IBIS/ISGRI time-resolved spectroscopy of the burst \citep{Gotz:2003} was
consistent with the overall hard-to-soft evolution observed by
BATSE in brighter GRBs \citep{Preece:1998,Ford:1995}. However, the
fluence of GRB 030131 was an order of magnitude smaller, indicating
that such spectral behavior also applied to fainter GRBs.
Clear evidence of such behavior was also reported for bursts studied
with BeppoSAX in the 2-700 keV energy range (e.g. \citep{Frontera:2000}).

\subsection{From extremely hard to soft bursts}
\subsubsection{GRB 030406}
ISGRI and PICsIT can act as a Compton telescope,
registering photons that are scattered in one and absorbed
in the other detector. The coincidence time window is
a parameter programmable on-board and the mode is
sensitive between 200 keV and $\sim 3$ MeV.
Thanks to the Compton mode data, INTEGRAL has been able to make
successful, independent burst localizations, as in the case of GRB 030406 (\citep{Marcinkowski:2003} and references therein). GRB 030406 was detected by SPI-ACS on-board INTEGRAL as well as by
Ulysses, Konus, and Mars Odyssey \citep{Hurley:2003}. It was
a long burst lasting $\sim 65$ s. Ulysses reported a fluence
of $1.3\times10^{-5}$ erg cm$^{-2}$ s$^{-1}$ (25-100 keV).
In the 50 keV - 3 MeV energy range, a fit to the spectrum of the 
burst peak (2.8 s around the maximum) with a broken power law, 
revealed that the peak was very hard. Below 400 keV, the $\nu F_{\nu}$ spectrum increased with index $\sim +3.5$, and above
this energy the index was still positive and equal to $\sim +0.3$ \citep{Marcinkowski:2003}. 
Compared to the distribution of the low and high
energy spectral indices of the sample of 156 bright BATSE bursts presented by \citep{Preece:2000},
both the low and the high energy spectral indices of GRB 030406 would lie in 
the upper tails of the corresponding distributions in this BATSE sample. Its
spectrum was thus in clear contradiction with
the synchrotron model of GRBs \citep{Katz:1994,Tavani:1995},
which predicts that there is a strict upper limit on the low energy
spectral index of $-2/3$ \citep{Preece:1998}. The low energy
spectral slope of the burst was instead consistent within the error bars with the jitter
radiation model of GRBs \citep{Medvedev:2000}. \citep{Ghirlanda:2003}
also showed that the low energy spectral slopes of several GRBs are indeed much larger than expected in the
synchrotron model ($\sim 0.5-1$).
Furthermore, GRB 030406 provided a hint for the
existence of bursts with the peak of $\nu F_{\nu}$ spectrum, $E_{peak}$, above the
distribution shown by \citep{Mallozzi:1995} which truncates at $\sim 1$ MeV for the bright BATSE bursts.
An $E_{peak}>1.1$ MeV, as estimated for GRB 030406 ($90\%$ lower limit), is also on the
high end of the Amati (and Ghirlanda) relation \citep{Amati:2002,Ghirlanda:2004}. 
According to such a relation, the estimated isotropic radiated energy is
larger than $(6-20)\times10^{53}$ ergs, where the uncertainty stems from
the inaccuracy of the Amati relation \citep{Marcinkowski:2003}. 
GRB 030406 thus showed that hard bursts are very energetic.
In summary, GRB 030406 confirmed that INTEGRAL in the
Compton mode can detect hard GRBs, with a localization accuracy
of a few degrees \citep{Marcinkowski:2003}. The GRB spectra can be studied in the
range from $\sim 200$ keV to $\sim 3$ MeV depending on the location of
the burst in the instrument coordinates.

\subsubsection{XRF 040812}
\begin{figure}
	\centering
		\includegraphics[width=0.45\textwidth]{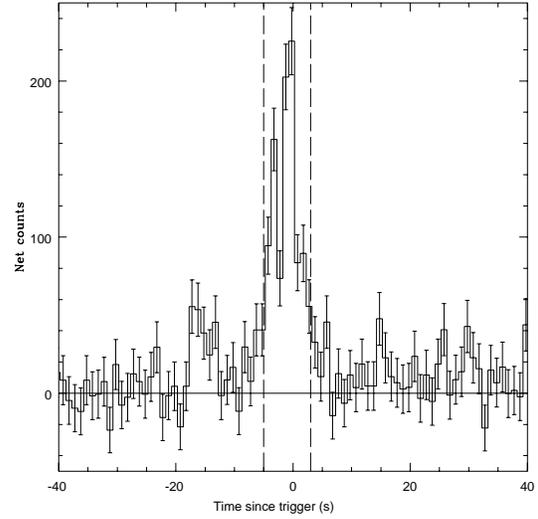}
		\caption{The IBIS/ISGRI 20 - 60 keV light curve of XRF 040812. 
\label{Fig1}}
\end{figure}

\begin{figure}
	\centering
		\includegraphics[width=0.45\textwidth]{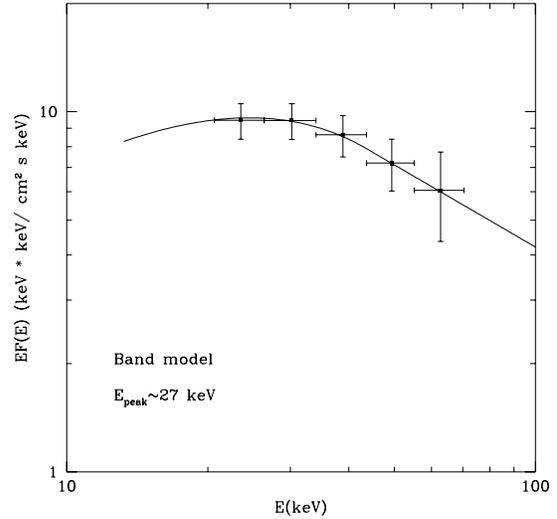}
		\caption{XRF 040812 spectrum: assuming a Band model \citep{Band:1993}, we get $\alpha\sim-1$ and $\sim -3$ for the low and high energy spectral indices of the photon spectrum, consistent with typical XRF values. The $\nu F_{\nu}$ peak energy is at $E_{peak} \sim 27$ keV.
\label{Fig2}}
\end{figure}

\begin{figure}
	\centering
		\includegraphics[width=0.45\textwidth]{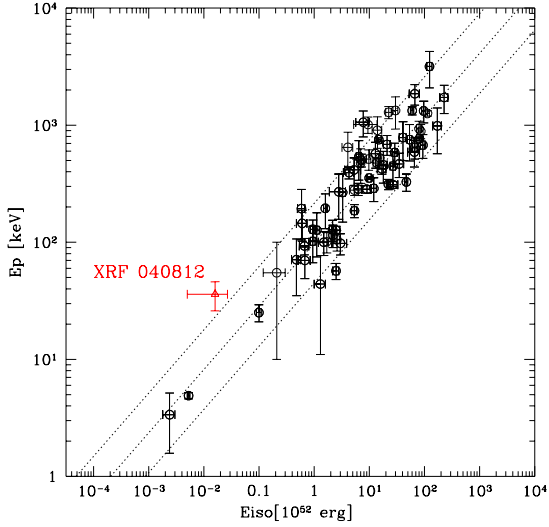}
		\caption{$E_{peak}-E_{iso}$ correlation for XRF 040812 compared with the 70 GRBs and XRFs with firm estimates of the redshift reported in \citep{Amati:2008}. XRF 040812 ranks  among XRRs rather than intrinsic XRFs.
\label{Fig3}}
\end{figure}

In addition to very hard bursts, INTEGRAL has also yielded 
interesting results on much softer GRBs, as for the case of the X-Ray Flash (XRF) 040812.
XRF 040812 was discovered by INTEGRAL on 2004 August 12.251 UT, 
with the IBIS/ISGRI instrument in the 15-200 keV band (\citep{GCN:2640} and Fig. \ref{Fig1}). A spectral analysis of the
burst revealed good agreement with a Band model \citep{Band:1993}, with low energy photon index $\alpha=-1$ and high energy photon index $\beta=-3$. The $\nu F_{\nu}$ spectrum peaked at an energy $E_{peak}\sim 27$ keV (see Fig. \ref{Fig2}), and the 20-200 keV flux was measured to be $F_{20-200 {\rm keV}}\sim 2.2\times10^{-8}$ ergs cm$^{-2}$ s$^{-1}$. Two epochs of 
Chandra observations (5 and 10 days after the burst) revealed the presence of a fading, 
uncatalogued source within the INTEGRAL error box \citep{GCN:2649,GCN:2656}. Optical observations
at the VLT started 0.7 days after the XRF and the monitoring continued until 220 days after \citep{Davanzo:2006},
when the level of the host galaxy emission was reached. These observations constrained the redshift to lie in the range $0.3<z<0.7$ \citep{Davanzo:2006}. Taking into account the cosmological redshift, and using INTEGRAL data, it was possible to establish that XRF 040812 was consistent with the $E_{peak}-E_{iso}$ relation, ranking among X-Ray Rich GRBs rather than an intrinsic XRFs (see Fig. \ref{Fig3} and Fig. 6 in \citep{Stratta:2007}).

\subsection{Peculiar spectral features: Fe-line and inverse Compton emission}
GRB 030227 was the first GRB for which the quick localization 
obtained with the IBAS has led to the discovery of X-ray and 
optical afterglows (see \citep{Mereghetti:2003} and references therein). 
The IBAS alert message with the preliminary
 position of the burst was delivered to the IBAS Team at 08:42:38 
 UT on 2003 February 27, only 35 s after the start of the burst (most of this delay was
due to buffering of the telemetry on board the satellite and to data 
transmission between the ground station and the ISDC). Unfortunately, 
the Internet message with these coordinates
could not be distributed in real time. In fact, the detection of 
GRB 030227 occurred during a calibration observation
of the Crab Nebula. Since the instrument configuration during these observations
caused some spurious IBAS triggers, the automatic delivery of the
alerts to external clients had been temporarily disabled. Nevertheless, this information
was distributed within less than 1 hr after the GRB (at 08:42:03
UT). GRB 030227 had a duration of about 20 s, thus belonging to
the class of long GRBs. Its peak flux and fluence, converted to
the BATSE 50-300 keV energy range, were 
$\sim 0.6$ photons cm$^{-2}$ s$^{-1}$ and $\sim 7\times10^{-7}$ ergs cm$^{-2}$, respectively \citep{Mereghetti:2003}. 
About three-quarters of the GRBs in the
fourth BATSE catalog \citep{Paciesas:1999} had a higher peak
flux, indicating that GRB 030227 was quite faint. The burst time-averaged 
spectrum was fit by a single power-law with photon index $1.9\pm0.3$. 
The 20-200 keV flux of the burst was measured to be $4.7\times10^{-8}$ ergs cm$^{-2}$ s$^{-1}$. 
Based on the background-subtracted count rates in the ranges 40-100 keV and 20-40 keV, both
SPI and IBIS showed evidence for a hard-to-soft evolution of the burst \citep{Mereghetti:2003}.

XMM-Newton observed the position of GRB 030227 for $\sim 13$ hrs, starting on February 27 at 16:58 UT, about 8 hrs after the event \citep{Mereghetti:2003}. 
An absorbed power-law fit to the afterglow spectrum gave an unacceptable fit. An acceptable one could be obtained
by fixing $N_{H}$ to the Galactic value and adding a redshifted neutral absorption, which yielded an estimated redshift of $z=3.9\pm0.3$ \citep{Mereghetti:2003}. 
The fit residuals suggested the presence of possible lines in the spectrum. Adding Gaussian lines
at different energies and with fixed widths, smaller than the
instrumental resolution, a possibly significant line was
found at an observed energy of $1.67\pm0.01$ keV \citep{Mereghetti:2003}. 
According to an F-test, such a line was significant at the $3.2\sigma$ level (to be considered with caution, see \citep{Protassov:2002}).
If interpreted as being due to Fe, as suggested for similar features observed
in other afterglows (see e.g. 
\citep{Piro:1999a,Piro:1999b,Antonelli:2000,Amati:2000,Piro:2000,Protassov:2000,Yoshida:2001}), the implied redshift
of $\sim 2.7-3$ (depending on the Fe ionization state) was
consistent with the value derived from the absorption (but other interpretations in terms of lighter elements were also possible \citep{Watson:2003}).

ToO observations were triggered starting at 5.3 hrs with different optical telescopes \citep{Castro-Tirado:2003}. 
The spectral flux distribution of the optical afterglow of
GRB 030227 on 28.24 UT Feb 2003, determined by means of the broad-band
photometric measurements obtained with the different telescopes, 
gave a NIR/optical spectral index of $\beta_{opt-NIR}=-1.25\pm0.14$, 
consistent with the spectral index for the unabsorbed
X-ray spectrum ($\beta_X = -0.94\pm0.05$, \citep{Mereghetti:2003}). 
However, the NIR/optical and X-ray spectra did not match each other's extrapolations, 
similarly to other previously and more recently observed GRBs \citep{Harrison:2001,intZand:2001,Corsi:2005}. 
Considerable extinction in the host galaxy (considering different extinction
laws) could not explain such an effect \citep{Castro-Tirado:2003}. 
Thus, it was suggested that in contrast
to the NIR/optical band, where synchrotron processes dominate,
an important contribution of inverse Compton
scattering was present in the X-ray spectrum, besides line emission \citep{Mereghetti:2003,Watson:2003}. 
This implied a lower limit on the density of the external
medium of $n>10$ cm$^{-3}$ \citep{Castro-Tirado:2003}.

\subsection{A low-luminosity, nearby GRB}
\begin{figure}
\includegraphics[width=0.45\textwidth,
height=0.25\textheight]{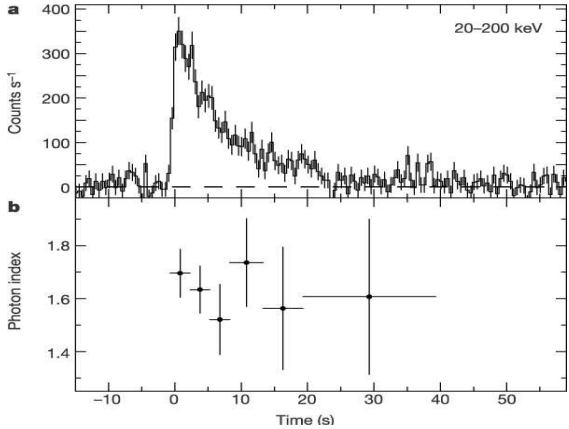}
\caption{\textbf{a)} Lightcurve of GRB\,031203 in the 20--200\,keV energy band in 0.5\,s time bins. \textbf{b)} The evolution of the photon index throughout the duration of the burst. There is no evidence for significant spectral evolution on short timescales. Credit: \citet{sazonov:2004}.}
\label{fig:031203}
\end{figure}

GRB\,031203 was detected by IBIS on December 3, 2003 at 22:01:28 UTC. The light curve in Fig.~\ref{fig:031203}\,a shows a single pulse with a duration of $\sim$40\,s.  The spectrum was also typical for a long-duration GRB and was well fit by a single power-law model that constrained $E_{peak}>190$\,keV \citep{sazonov:2004}. GRB\,031203 is the fourth nearest GRB detected to date at a redshift of $z=0.1055$ and is notable for its unambiguous spatial and temporal association with the supernova SN\,2003lw. It is a confirmed low-luminosity GRB with an isotropic energy of $\sim4\times10^{49}$\,erg (20--200\,keV). GRB\,980425, with an isotropic energy release of $\sim10^{48}$\,erg was the first such underluminous GRB to be detected and was associated with the nearby ($z=0.0085$) SN\,1998bw. Both GRB\,980425 and GRB\,031203 violate the Amati correlation between isotropic energy, $E_{iso}$ and peak energy, $E_{peak}$, that would predict $E_{peak}<10$\,keV \citep{ghisellini:2006}. These GRBs are clearly sub-energetic in the $\gamma$-ray band, and their proximity (and hence implied abundance) makes it of great interest to understand their origin and relation to the more distant cosmological GRBs. 

\subsection{Polarisation studies}
Linear polarisation in GRB prompt emission is an important 
diagnostic with the potential to significantly constrain GRB emission models 
(see e.g. \citep{Toma:2008} and references therein for a recent discussion of this topic). 
GRB 041219a is the most intense burst localised by INTEGRAL with a fluence 
of $5.7\times10^{-4}$ ergs cm$^{-2}$ over the energy range 20 keV - 8 MeV \citep{McGlynn:2007},
and is an ideal candidate for such a study. 
GRB 041219a was detected by IBAS and ISGRI at 01:42:18 UTC on
December 19th 2004 \citep{Gotz:2004}. The burst was in the fully 
coded field of view of both ISGRI and SPI and was $3^{\circ}$ off the X-axis, and $155^{\circ}$ in azimuth from the
Y-axis. It was also detected by the Burst Alert Telescope (BAT)
on-board the \textit{Swift} satellite \citep{Barthelmy:2004}. A $T_{90}$ value of 186 s 
($\sim 20$ keV -8 MeV) was determined from the SPI light curve. The temporal structure 
of the burst was unusual, with an initial weak precursor pulse followed by a long quiescent time interval
and the main emission beginning at $\sim 250$ s post trigger \citep{MacBreen:2006}. 
Although the SPI light curve is quiescent from $\sim 7-200$ s, emission was 
detected in the BAT light curve, particularly in the lower energy channels 15-25 keV and 25-
50 keV \citep{Fenimore:2004}. In addition, a spectrally soft pulse was detected 
in the All Sky Monitor (ASM) \citep{Levine:1996}, beginning at 80 s in the quiescent period of the SPI observation \citep{Levine:2004}. 

Polarisation can be measured by INTEGRAL/SPI using
multiple events scattered into adjacent detectors because the Compton 
scatter angle depends on the polarisation of the incoming
photon. For the case of GRB 041219a, \citep{Kalemici:2007} 
found that the distribution of azimuthal scattering angles
was better represented by a polarized source
compared to a non-polarized source, but with low statistical significance.
Using multiple-detector coincidence events in the 100-350 keV energy band, 
such analysis yielded a polarization fraction $(98\pm33)\%$ \citep{Kalemici:2007}, which 
statistically did not allow the authors to claim a polarization detection from this source. 
Moreover, different event selection criteria lead to even less significant polarization fractions, 
e.g., lower polarization fractions were obtained when higher energies were 
included in the analysis \citep{Kalemici:2007}, and the possibility that the
measured modulation was dominated by instrumental systematics could not be ruled out \citep{Kalemici:2007}.

\begin{figure}
\includegraphics[width=0.45\textwidth,
height=0.25\textheight]{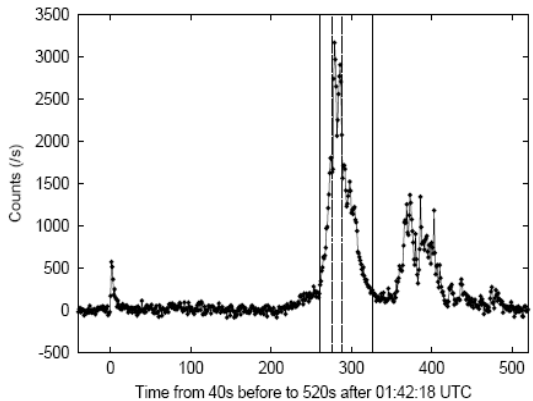}
\caption{Background-subtracted single event lightcurve of
GRB 041219a, summed over all SPI detectors in the energy
range 20 keV - 8 MeV. The vertical solid lines mark the start
and end of the 66 second emission phase ($T_0$ = 261 s to $T_0$ =
327 s). The vertical dashed lines mark the start and end of the
brightest 12 seconds of the burst ($T_0$ = 276 s to $T_0$ = 288 s). $T_0$
is the IBAS trigger time (01:42:18UTC). Credit: McGlynn et al. 2007 \citep{McGlynn:2007}. \label{pol}}
\end{figure}

\citep{McGlynn:2007} have also carried out polarisation studies on GRB 041219a, comparing the observed data
to various combinations of simulated polarised and unpolarised
data. The analysis was carried out dividing the scatter pairs into 6 
different directions (0 - 300 degrees) and in the energy
ranges 100 - 350 keV and 100 - 500 keV. The degree of linear polarisation in the 
GRB brightest pulse of duration 66 s (see Fig. \ref{pol}) was found to be $(63^{+31}_{-30})\%$
at an angle of $(70^{+14}_{-11})$ degrees in the 100 - 350 keV energy range. 
The degree of polarisation was also constrained in the brightest
12 s of the GRB (see Fig. \ref{pol}) and a polarisation fraction of $(96^{+39}_{-40})\%$ 
at an angle of $(60^{+12}_{-14})$ degrees was determined over the same energy range.
However, despite extensive analysis and simulations, 
a systematic effect that could mimic the weak polarisation signal could not
be definitively excluded. Over several energy ranges and time intervals, the results
by \citep{McGlynn:2007} were consistent with a polarisation signal of
about $60\%$ but at a low level of significance ($\sim 2 \sigma$). 
By fitting the azimuthal scatter angle distribution of the observed
data over the 6 directions, \citep{McGlynn:2007} obtained results consistent with
\citep{Kalemici:2007} in both magnitude and direction, within
the limits given by the large error bars. Also, \citep{McGlynn:2007} agreed with the conclusions
of \citep{Kalemici:2007} that the possibility of
instrumental systematics dominating the measured effect cannot be excluded.

More recently, \citep{Gotz:2009} reported on the polarization measurement of the prompt
emission of GRB 041219a, performed using IBIS. Being composed of the two position sensitive 
detector layers ISGRI and PICsIT, IBIS can be used
as a Compton polarimeter \citep{Lei:1997}, thanks to the
polarization dependency of the differential cross section for Compton scattering.
\citep{Gotz:2009} analyzed the different portions of the GRB 040219a prompt light curve, focusing
on the brightest parts. More specifically, they analyzed the entire first and
second peak, and then performed a time-resolved analysis composed of
36 intervals lasting 10 s, each one overlapping for 5 s with
the previous one, over the whole duration of the GRB starting at 01:46:22 UT until 01:49:22 UT. 
No polarization signal could be found integrating over the whole first peak, with an upper limit of
$4\%$. On the other hand, a modulated signal was seen in the second peak, corresponding to a polarization
fraction of $43\%\pm25\%$. Integrating over smaller portions of the GRB, highly polarized
signals were measured. Results by \citep{Gotz:2009} were consistent with those by \citep{McGlynn:2007}
for the 12 s interval on the brightest part of the GRB. However, \citep{Gotz:2009} did not confirm 
the result by \citep{McGlynn:2007} on the broader 66 s time interval, where they did not
detect any polarized signal. 

\citep{Gotz:2009} have discussed their results in the light of different emission models for GRB prompt emission: (i) Synchrotron emission from shock-accelerated electrons in
a relativistic jet with an ordered magnetic field contained
in the plane perpendicular to the jet velocity \citep{Granot:2003a,Granot:2003b,Nakar:2003}; (ii) Synchrotron emission from a purely electromagnetic outflow \citep{Lyutikov:2003}; (iii) Synchrotron emission from shock-accelerated electrons in
a relativistic jet with a random field generated at the shock and contained in the plane perpendicular to the jet velocity \citep{Ghisellini:1999,Waxman:2003}; (iv) Inverse Compton emission from relativistic electrons in
a jet propagating within a photon field (``Compton drag'' model, \citep{Lazzati:2004}).
Given that for the case of GRB 041219a the polarization level was varying on short timescales (reaching high
values), that the polarization angle was varying as well, and that the time-averaged value over longer intervals showed reduced polarization, \citep{Gotz:2009} concluded that, despite being consistent with more than one model, their results for this GRB favor synchrotron radiation from a relativistic outflow with a magnetic field, which is coherent on an angular size comparable with the angular size of the emitting region ($\sim 1/\Gamma$, scenario (i)).

\section{Global Properties of GRBs observed by INTEGRAL}

\subsection{Duration, spectral index and peak flux distributions}

\begin{figure}
\includegraphics[width=0.45\textwidth,
height=0.25\textheight]{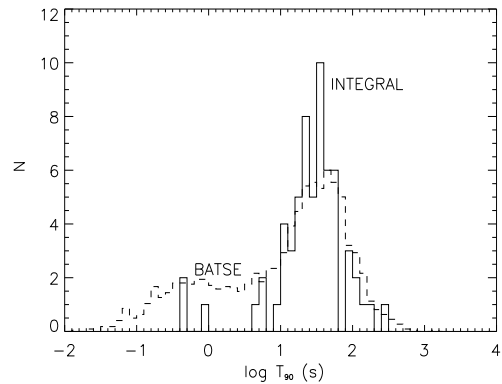}
\caption{$T_{90}$ distribution of INTEGRAL GRBs in comparison to that of BATSE (dashed line). The BATSE distribution is normalised to the INTEGRAL distribution for clarity.}
\label{fig:int_t90}
\end{figure}

The $T_{90}$ distribution of INTEGRAL GRBs, measured in the 20--200\,keV energy range, is shown in Fig.~\ref{fig:int_t90} and compared with the bimodal distribution obtained for BATSE GRBs \citep{kouveliotou:1993}. There is reasonable agreement between the two distributions, especially when the small number of INTEGRAL GRBs is taken into account. 

\begin{figure}
\includegraphics[width=0.45\textwidth,
height=0.25\textheight]{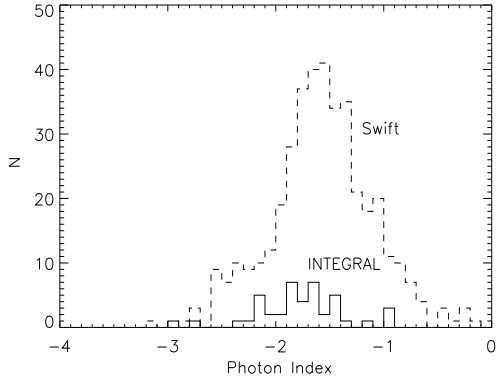}
\caption{Power-law photon index distributions for INTEGRAL (solid line) and Swift (dashed line) The Swift data for 395 GRBs is taken from http://swift.gsfc.nasa.gov/docs/swift/archive/grb\_table.html.}
\label{fig:phot_ind}
\end{figure}

The distribution of photon indices is shown in Fig.~\ref{fig:phot_ind} for INTEGRAL and Swift GRBs for which a power-law model was fit to the spectral data in the 20--200\,keV and 15--150\,keV energy ranges, respectively. In comparison to Swift, INTEGRAL detects proportionally more soft GRBs with steeper power-law photon indices.

\begin{figure}
\includegraphics[width=0.45\textwidth,
height=0.25\textheight]{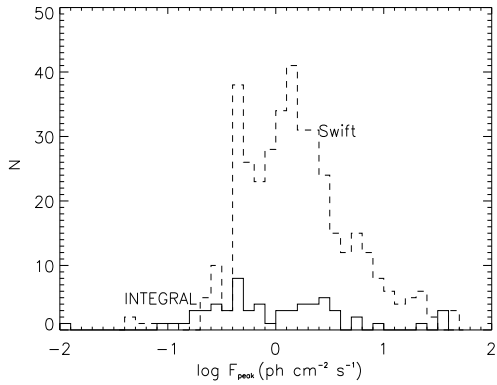}
\caption{Peak flux distribution for GRBs detected by INTEGRAL (20--200\,keV, solid line) and Swift (15--150\,keV, dashed line). The Swift data for 388 GRBs is taken from http://swift.gsfc.nasa.gov/docs/swift/archive/grb\_table.html.}
\label{fig:fpeak}
\end{figure}

Fig.~\ref{fig:fpeak} compares the peak flux distribution of the GRBs observed by IBIS to that observed by the BAT instrument on Swift. IBIS detects proportionally more weak GRBs than Swift because of its better sensitivity within a field of view that is smaller by a factor of $\sim$12. 

\subsection{Spectral lags}

There are a number of redshift indicators that can be used for GRBs without spectroscopic redshift measurements. One such indicator is the spectral lag which combines the spectral and temporal properties of the prompt GRB emission. The spectral lag is a measure of the time delay between GRB emission
 in a high-energy $\gamma$-ray band relative to the arrival of photons in a low-energy
 band. The typical lag values 
measured for long-duration GRBs detected by the Burst and Transient Source Experiment (BATSE) between the
25--50\,keV and 100--300\,keV channels
concentrate at $\sim100$\,ms~\citep{norris:2000}.  An
anti-correlation between spectral lag and isotropic
peak luminosity was first observed by \citet{norris:2000}, using 6 BATSE bursts
with measured redshifts. However, there exist notable
outliers, in particular the ultra-low luminosity
bursts GRB\,980425, GRB\,031203 and GRB\,060218, associated with the supernovae
SN\,1998bw, SN\,2003lw and SN\,2006aj, respectively.

In order to measure the lags of INTEGRAL GRBs,
background-subtracted lightcurves were extracted in 25--50\,keV and 50--300\,keV energy bands.
The lag, $\tau$, between two energy
channels was determined by computing the
cross-correlation function (CCF) between the two lightcurves
as a function of temporal lag as described by
\citet{band:1997} and ~\citet{norris:2000}. Assuming the time profiles  in both energy
channels display sufficient similarity, the peak in the CCF
then corresponds to the time lag of the GRB between the two
energy channels in  question.  

\begin{figure}
\includegraphics[width=0.45\textwidth,
height=0.25\textheight]{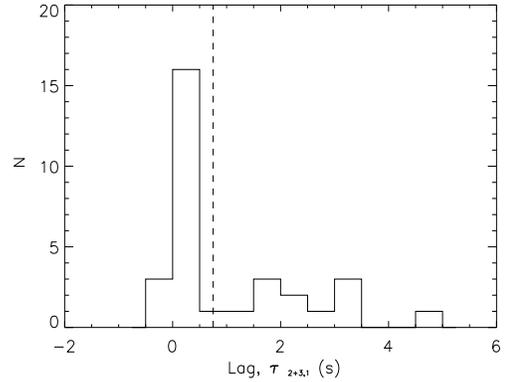}
\caption{Spectral lag distribution for the 30 INTEGRAL GRBs for which a lag could be measured between 25--50\,keV and 50--300\,keV. The distribution is separated by the dashed line into short-lag and long-lag GRBs at $\tau$=0.75\,s.}
\label{fig:lag_hist}
\end{figure}

The number distribution of 
spectral lags is given in Fig.~\ref{fig:lag_hist} for the 30 long-duration
GRBs with a measured lag between  25--50\,keV and 50--300\,keV. No statistically
significant negative spectral lags are found. A long tail extending to $\sim5$\,s is observed in the lag
distribution in Fig.~\ref{fig:lag_hist} and a clear separation between
short and long lag is drawn at $\tau\sim0.75$\,s. Thus, long-lag bursts have $\tau>0.75$\,s and those with $\tau<0.75$\,s are
referred to as short-lag GRBs. There are 12 long-lag GRBs in the INTEGRAL sample.

\begin{figure}
\includegraphics[width=0.35\textwidth,
height=0.35\textheight,angle=270]{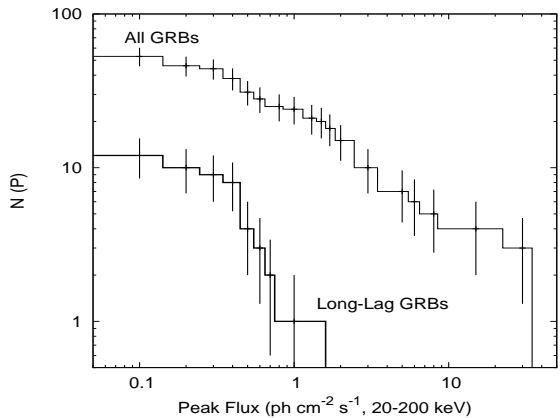}
\caption{The cumulative log N-log P distribution of the GRBs detected by IBIS, with peak flux, P, measured between 20--200\,keV. The small subset of 12 long-lag GRBs is shown separately.}
\label{fig:lognlogs}
\end{figure}

The logN-logP distribution is given in Fig.~\ref{fig:lognlogs} for all IBIS GRBs and separately for the subset of 12 long-lag GRBs. The distribution is biased by the lower sensitivity of IBIS at large off-axis angles. However, the long-lag GRBs appear to form a separate population at low peak fluxes.  

\subsection{Spatial distribution in Supergalactic coordinates}

\begin{figure}
\includegraphics[width=0.45\textwidth,height=0.25\textheight]{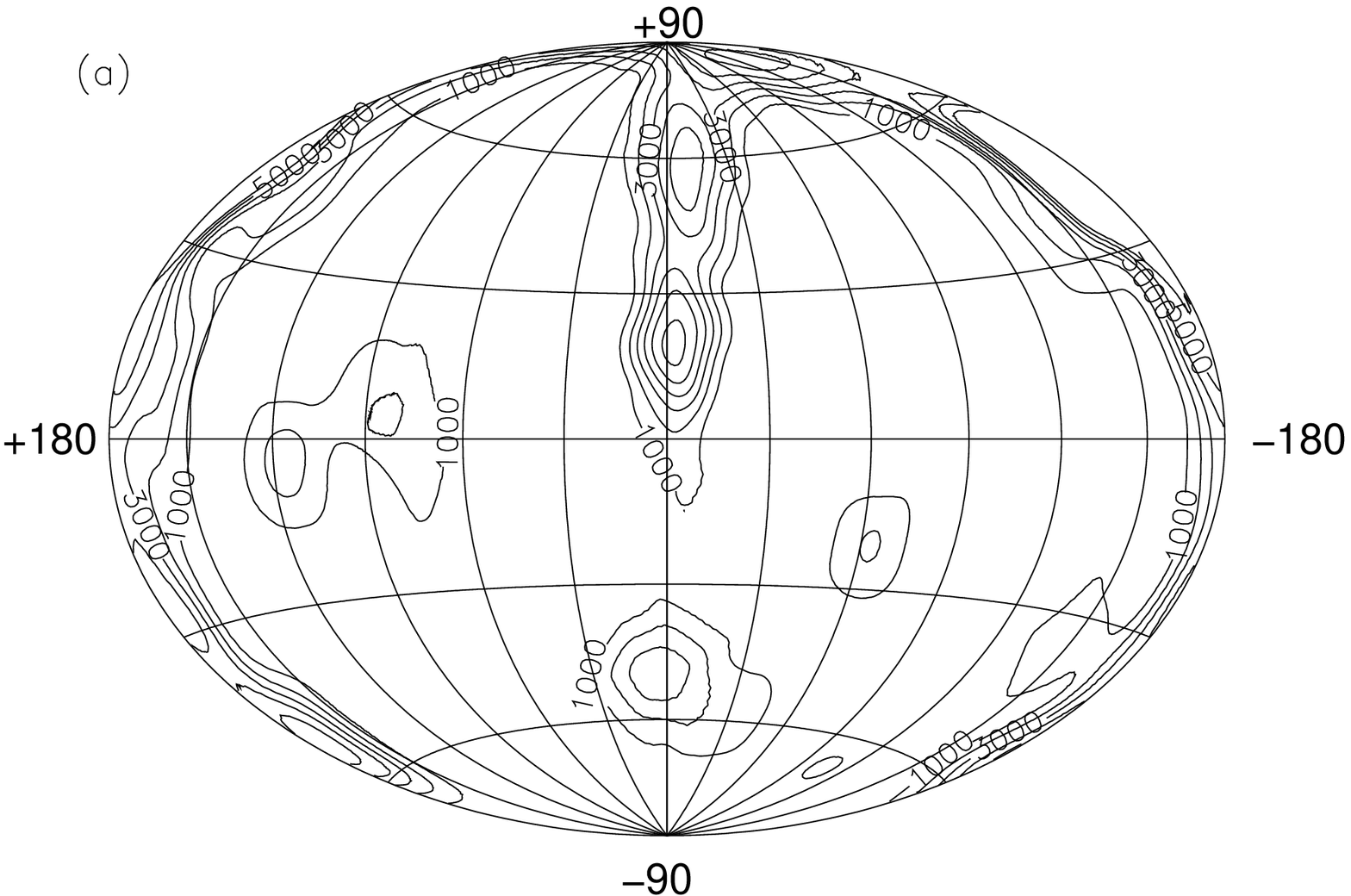}
\includegraphics[width=0.45\textwidth, height=0.25\textheight]{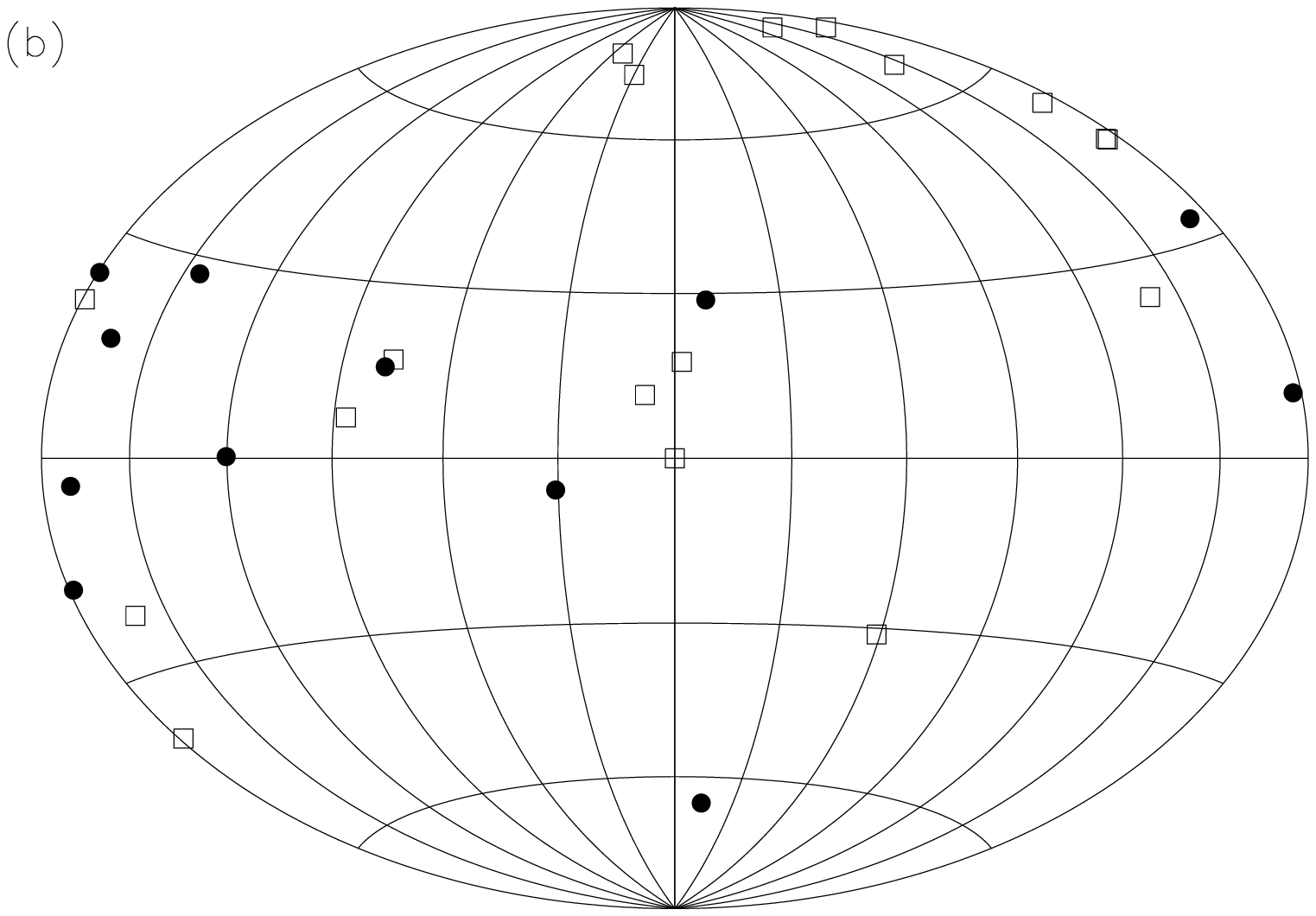}
\caption{\textbf{a} INTEGRAL exposure map in supergalactic coordinates (contours in units of kiloseconds). \textbf{b} The distribution of INTEGRAL GRBs in supergalactic coordinates; the open squares represent short-lag GRBs ($\tau<0.75$\,s) and filled circles those GRBs with long lags ($\tau>0.75$\,s).}
\label{fig:lags_sg}
\end{figure}

The INTEGRAL exposure map and distribution of GRBs in supergalactic coordinates is shown in
Fig.~\ref{fig:lags_sg}. All of the \textit{INTEGRAL} GRBs are divided almost equally between the
half of the sky above and below $\pm30^{\circ}$, in agreement with the
exposure map which has $\sim52$\% of the exposure time
within $\pm30^{\circ}$ of the supergalactic plane. However, 10 of the 12 long-lag GRBs are concentrated at supergalactic
latitudes between $\pm30^{\circ}$. The quadrupole moment~\citep{hartmann:1996} has a value of $Q=-0.166\pm0.086$ for the long-lag GRBs and indicates an anisotropy in the distribution of these GRBs with respect to the supergalactic plane. This result leads us to 
conclude that long-lag GRBs may trace the features of the nearby large-scale structure of the Universe and is a further indication that most long-lag GRBs are nearby and have low luminosity. 

A nearby population of long-lag, low-luminosity GRBs has previously been
proposed based on the detections of  GRB\,980425 and
XRF\,060218~\citep[e.g.][]{guetta:2004}. The distance to the long-lag GRBs can also be constrained by association and comparison with the two low luminosity bursts GRB\,980425 ($\tau\sim2.8\,\rm{s}$) and XRF\,060218 ($\tau\sim60\,\rm{s}$) which would have been detected in the fully-coded field of view of IBIS to 135\,Mpc and 290\,Mpc, respectively.  The association of the long-lag GRBs with known low luminosity GRBs and with the supergalactic plane implies that they are at similar distances. 
It has also been pointed out that weak BATSE GRBs appear to be correlated with galaxies out to distances of $\sim$155\,Mpc~\citep{chapman:2007}.

GRBs have a long lag when a typical value of 0.1\,s is redshifted by a large factor or alternatively it is an intrinsic property of a low-luminosity GRB such as GRB\,980425 and XRF\,060218. The rate of $z>5$ GRBs in IBIS has been modelled~\citep[e.g.][]{gorosabel:2004} and is unlikely to be more than 1 or 2 GRBs in 5 years of observations. We evaluate the rate of such GRBs over the whole sky using the 8 long-lag \textit{INTEGRAL} GRBs in the partially-coded field of view of IBIS at 50\% coding (0.1\,sr) over an exposure time of 4 years, adopting a distance of 250 Mpc and assuming that 2 of the 8 GRBs are at high redshift. We obtain $2640\,\rm{Gpc}^{-3}\,\rm{yr}^{-1}$, in which the major uncertainty is the distance. The rate of low-luminosity GRBs at the adopted distance of 250 Mpc exceeds the upper limit of $<300\,\rm{Gpc}^{-3}\,\rm{yr}^{-1}$ of Type Ib/c SN producing GRBs, which was derived assuming that all low luminosity GRBs would produce a SN and be as radio bright as the SN GRBs~\citep{soderberg:2006}. However, the low luminosity GRB\,060605 has no associated SN to faint limits and is evidence for a quiet end for some massive stars~\citep[e.g.][]{fynbo:2006}.  

The association of low luminosity GRBs with the supergalactic plane is not proof that they are associated with clusters of galaxies but indicates that clusters may play a role. It is interesting to note that the rate of Type Ia SNe is higher in elliptical galaxies in clusters than in field ellipticals by a factor of $\sim$3~\citep{mannucci:2007}.  This effect is due to galaxy-galaxy interactions which can affect the evolution and properties of binary systems. In this case, there should also be an increase in the merger rate of white dwarfs or a white dwarf with a neutron star or black hole.  A merger involving a white dwarf~\citep[e.g.][]{king:2007} should produce a long GRB that is likely to be fainter than the formation of a black hole in cosmological GRBs.  There will be no supernova in the merger of a white dwarf with a neutron star or black hole, and probably a faint afterglow. 

\section{Conclusion}
INTEGRAL is an Observatory type Gamma-ray mission with two main instruments
optimized for high resolution $\gamma$-ray imaging and spectroscopy in the energy
range from 15 keV up to 10 MeV.

The observatory, even if  not fully optimized for GRB science (mainly
due to on-board limited resources in term of TLM, buffering memory and
computing power etc.) features a substantially large field of view of the
main instruments ($> 1000$ squared degrees), real time photon by photon
transmission to ground, and the IBAS system that distributes the position of the GRB very effectively, with an accuracy of the order of 1
to 3 arcmin a few seconds after the onset of the burst.

From its launch in October 2002 to date, INTEGRAL has detected and
localised 65 GRBs. They have provided significant insight into the
prompt $\gamma-$ray emission in the hard-X/soft-$\gamma$-ray energy range, i.e.
where GRBs emit a large fraction of their energy. Furthermore, INTEGRAL
provides unique sensitivity in the MeV range via the IBIS Compton mode, 
an energy region which was not sensitively covered by other operative satellites until the recent
launches of AGILE and FERMI. Also, for the brightest bursts, 
both IBIS and SPI have independent, moderate, polarization
capabilities.

Due to the limited FoV the overall rate of accurately positioned GRBs is
limited to about 0.8 GRBs per month, with real time localisations
enabling multi-wavelength observations to be carried out
by other space-based missions and ground-based telescopes. On the other
hand, when compared with SWIFT, a GRB oriented mission, INTEGRAL detects
proportionally more faint GRBs due to its narrower FoV that limits the count rate due X-Ray Diffuse Background, which is the limiting factor to sensitivity in the lower operative energy range ($ E < 100 - 150$ keV).
The sensitivity of IBIS is such that it can detect very faint GRBs,
allowing the investigation of the population of low-luminosity GRBs with long lags which appears to be a low-luminosity population distinct from the
high-luminosity one and is inferred to be local. Finally, the all-sky
rate of GRBs with a flux above  0.15 ph cm$^{-2}$ s$^{-1}$ is 1400 yr$^{-1}$, based
on the bursts detected in the IBIS fully coded field of view.

After almost 7 years of successful  operation in orbit, the European Space
Agency is in the process of extending INTEGRAL operations beyond 2012.
INTEGRAL has all the redundant system still unused, all of the on board
systems perfectly operative, including the two main instruments and
monitors. The mission's extension will enable us to collect several tens of new
GRBs, in turn allowing us to confirm with higher statistical confidence the
local nature of the long lag ones.

We believe it is mandatory to maintain INTEGRAL in operation as an
important space observatory open to the community at large,
operating in full synergy with XMM, CHANDRA, Suzaku, RXTE and SWIFT for
the low energy part of the X-ray domain, and AGILE and FERMI for the
$\gamma$-ray range.

The impressive space satellite fleet now available to the scientific
community has resulted in the present golden age for high energy astrophysical
science, while anticipating the the new generation of $\gamma$-ray flight missions such as EXIST, etc.


\begin{theacknowledgments}
PU and AB acknowledge financial contribution from ASI-INAF  contract I/008/07/0. AC is grateful to the Italian L'Oreal-UNESCO program ``For Women in Science'' and acknowledges the support of ASI-INAF contract I/088/06/0.
\end{theacknowledgments}


\bibliographystyle{aipproc}   

\bibliography{Proc_AIP_Ubertini_Pietro_IR_P}

\end{document}